\documentclass[aps,pra,twocolumn,showpacs,superscriptaddress,floatfix]{revtex4-1}
\usepackage[T1]{fontenc}
\usepackage[latin9]{inputenc}
\usepackage{float}
\usepackage{amsmath}
\usepackage{amssymb}
\usepackage{nicefrac}
\usepackage{graphicx}
\usepackage{esint}
\usepackage{babel}

\usepackage{physics}

\def\kr{k_{\mathrm R}}                             
\def\Er{E_{\mathrm R}}                             

\def\ex{\mathbf{e}_x}

\usepackage{xcolor}
\usepackage[normalem]{ulem}

\begin{document}
\title{Topological charge pumping with subwavelength Raman lattices}

\author{D.~Burba}
\affiliation{Institute of Theoretical Physics and Astronomy, Vilnius University, A. Go\v{s}tauto 12, Vilnius LT-01108, Lithuania}

\author{M. Ra\v{c}i\={u}nas}
\affiliation{Institute of Theoretical Physics and Astronomy, Vilnius University, A. Go\v{s}tauto 12, Vilnius LT-01108, Lithuania}

\author{I.~B.~Spielman}
\email{spielman@nist.gov}
\homepage{http://ultracold.jqi.umd.edu}
\affiliation{Joint Quantum Institute, University of Maryland, College Park, Maryland 20742-4111, 20742, USA}
\affiliation{National Institute of Standards and Technology, Gaithersburg, Maryland 20899, USA}

\author{G.~Juzeli\={u}nas}
\email{gediminas.juzeliunas@tfai.vu.lt}
\affiliation{Institute of Theoretical Physics and Astronomy, Vilnius University, A. Go\v{s}tauto 12, Vilnius LT-01108, Lithuania}

\date{\today}

\begin{abstract}
Recent experiments demonstrated deeply subwavelength lattices using atoms with $N$ internal states Raman-coupled with lasers of wavelength $\lambda$.
The resulting unit cell was $\lambda/2N$ in extent, an $N$-fold reduction compared to the usual $\lambda/2$ periodicity of an optical lattice.
For resonant Raman coupling, this lattice consists of $N$ independent sinusoidal potentials (with period $\lambda/2$) displaced by $\lambda/2N$ from each other.
We show that detuning from Raman resonance induces tunneling between these potentials.
Periodically modulating the detuning couples the $s$- and $p$-bands of the potentials, creating a pair of coupled subwavelength Rice--Mele chains.
This operates as a novel topological charge pump that counter-intuitively can give half the displacement per pump cycle of each individual Rice--Mele chain separately.
We analytically describe this behavior in terms of infinite-system Chern numbers, and numerically identify the associated finite-system edge states.
\end{abstract}

\maketitle

\section{Introduction}

The behavior of one-dimensional (1D) systems is frequently tractable by analytic and numerical methods, often making them ideal prototypes for understanding phenomena that are intractable in higher dimensions. 
Even non-interacting systems such as those described by the Rice--Mele (RM) model~\cite{RiceMeleOriginal} can have non-trivial topology manifesting as protected edge states and quantized topological charge pumping~\cite{Thouless1983}.
Here we focus on a recently developed 1D subwavelength lattice for ultracold atoms built from $N$ Raman-coupled internal states~\cite{Anderson2020, Zhou2022PRXQ} and show that adding temporal modulation to the detuning away from Raman resonance can drive transitions between the $s$- and $p$-band Wannier states in adjacent lattice sites.
In the tight-binding limit, this gives rise to a pair of coupled RM chains with new regimes of topological charge pumping as well as topologically protected edge states.

\begin{figure}[tbh!]
\includegraphics{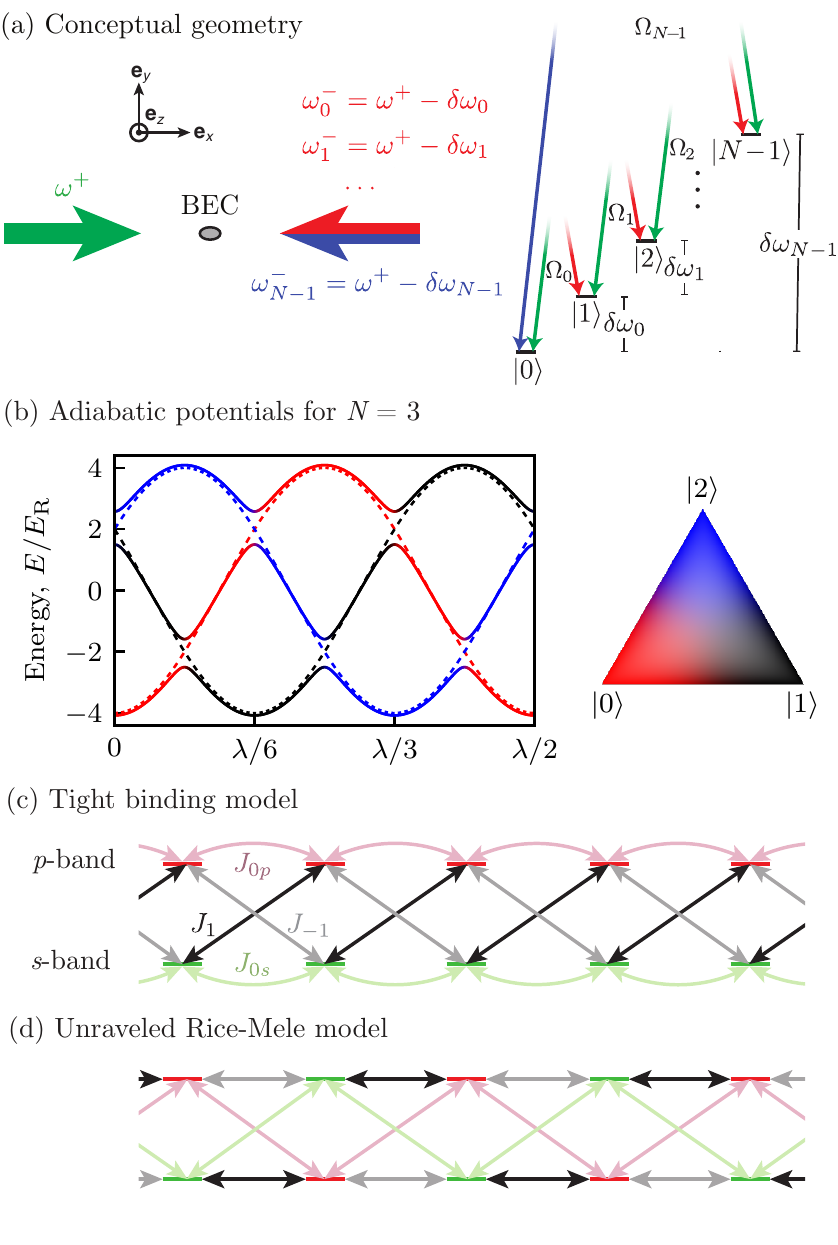}
\caption{Lattice concept.
(a) Experimental geometry with a single frequency Raman beam traveling along $\ex$ and $N$ Raman laser beams sharing the same spatial mode traveling along $-\ex$.
The level diagram for cyclic coupling is depicted on the right.
(b) Dressed state energies for $N=3$ and $\Omega_0 = \Omega_1 = \Omega_2 = 1\Er$. The dashed curves are computed for zero detuning, whereas the solid ones are calculated for a detuning described by Eq.~\eqref{eq:delta_j-oscil-detun} with $l=1$ and $\delta=0.5\Er$.
All curves are colored according to ternary plot on the right, marking the occupation probabilities in the three dressed states (not the bare internal atomic states) obtained by diagonalizing Eq.~\eqref{eq:H_0}. 
(c) Resonant driving gives nearest neighbor coupling $J_{\pm1}$ between the $s$- and $p$-bands.
Coupling within bands is induced by a static detuning with matrix elements $J_{0s}$ and $J_{0p}$. 
(d) The same lattice unraveled into coupled RM chains.
}
\label{fig:setup}
\end{figure}

Conventional optical lattices for ultracold atoms rely on the ac Stark shift to produce potentials proportional to the local optical intensity.
As a result, the lattice period can never be be smaller than half the optical wavelength $\lambda$.
Recently two techniques have emerged to create deeply sub-wavelength lattices~\cite{Wang2018,Anderson2020,Zhou2022PRXQ,Tsui2020}, both can be understood in terms of ``dressed states'' created by coupling internal atomic states with one- or two-photon optical fields~\cite{Dum1996,Juzeliunas2006,Campbell2011,Jendrzejewski2016,Lacki2016, Gvozdiovas2021,Li2022}.
Here we consider the scheme depicted in Fig.~\ref{fig:setup}(a) relying on sequentially coupling $N$ internal atomic states using two photon Raman transitions.
For resonant couplings of equal strengths, this results in independent adiabatic potentials for each of the $N$ dressed states, displaced by $\lambda/2N$ from each other as shown by the dashed curves in Fig.~\ref{fig:setup}(b).

This idealized situation is disturbed by imbalancing the coupling strengths, as studied in Ref.~\cite{Anderson2020}, or by detuning one or more of the transitions from resonance; the latter situation is plotted in Fig.~\ref{fig:setup}(b).
The addition of such perturbations makes evident the $\lambda/(2N)$ periodicity of the adiabatic potential, giving rise to nearest-neighbor (NN) tunneling between sites spaced by a single reduced unit cell.
This induced tunneling is generally much stronger than the natural $N$'th neighbor tunneling of the undisturbed lattice.

Here we focus on the effects of an additional time-modulated detuning which gives rise to an effective tunneling matrix element between $s$- and $p$-band Wannier states spaced by $\pm\lambda/(2N)$, leading to a novel subwavelength optical lattice.
In this lattice the proximity between adjacent sites allows the modulation induced matrix element to be comparable or larger than that of the NN tunneling induced by static detuning.
Fig.~\ref{fig:setup}(c) shows the resulting lattice geometry arising from this description, and (d) unwraps this into a pair of coupled Rice-Mele (RM) chains described by a highly tunable two-leg ladder Hamiltonian with novel topological properties that are the focus of this manuscript.

We study the topological aspects of this lattice both by considering adiabatic pumping and in terms of edge states. 
In the former case we show that the added inter-chain tunneling enables simple pumping trajectories giving per-cycle displacements of 0, 1 or 2 unit cells; by contrast only displacements in units of 2 sites are possible for the uncoupled RM chains.

This manuscript is organized as following.
In Sec.~\ref{sec:Hamiltonian} we formally derive the subwavelength Hamiltonian described above.
Section~\ref{sec:Symmetries} focuses on the subwavelength symmetry operations and solves the resulting band structure problem.
In Sec.~\ref{sec:TightBinding} we obtain a tight binding description of this lattice in terms of localized $s$- and $p$-band Wannier orbitals.
The band-changing tunneling induced by time-dependent detuning is derived in Sec.~\ref{sec:TimeDependent}.
Section~\ref{sec:Pumping} discusses the novel regimes of of topological pumping in the ladder.
The regimes of topological edge states are discussed in Sec.~\ref{sec:EdgeStates}.
Finally in Sec.~\ref{sec:Discussion} we expound on the implications of this work and conclude.

\section{Hamiltonian}\label{sec:Hamiltonian}

\subsection{Physical geometry}
\label{Physical geometry}

As illustrated in Fig.~\ref{fig:setup}(a), we consider an ensemble of ultracold atoms with $N$ internal atomic ground or metastable states $\ket{j}$ with $j=0,1,\ldots N-1$.
These states have nominal energies $\hbar\omega_j$, giving frequency differences $\delta\omega_j = \omega_{j+1} - \omega_j$, where here and below we adopt a periodic labeling scheme for which the labels $j$ and $j+N$ are equivalent; for example, this implies $\ket{j} = \ket{j+N}$ and $\omega_{j}=\omega_{j+N}$. 
Notice that for the specific energies depicted in (a), the state vector $\ket{N-1}$ has the largest energy and $\ket{0}$ has the smallest energy, making their frequency difference $\delta\omega_{N-1} = \omega_{0} - \omega_{N-1}$ negative. 

The atoms are illuminated by the pair of counterpropagating laser beams depicted in Fig.~\ref{fig:setup}(a) with wavelength $\lambda$, defining the single photon recoil momentum $\hbar\kr=2\pi\hbar/\lambda$ and energy $\Er=\hbar^2 \kr^2 / 2m$ for atoms of mass $m$.
The right going beam (green arrow) has angular frequency $\omega^+$ while the left going beam (red/blue arrow) has angular frequencies $\omega_j^- = \omega_+-\delta\omega_j$.
These lasers drive two-photon Raman transitions that cyclically couple the internal atomic states; each transition from $\ket{j+1}$ to $\ket{j}$ is characterized by an independent coupling strength $\Omega_{j}$.
The overall transition amplitude $-\Omega_{j}{\rm e}^{-{\rm i}2\kr x}$ includes a phase factor accounting for the two-photon recoil momentum $2 \hbar \kr$ imparted by the counter propagating lasers.
The resulting light-matter interaction is described by
\begin{equation}
\hat{V}\left(x\right)=-\sum_{j=0}^{N-1}\Omega_{j}{\rm e}^{-{\rm i} 2\kr x}|j\rangle\langle j+1|+{\rm H.\,c.},\label{eq:V-original}
\end{equation}
where a hat signifies an operator that acts on the internal atomic states, and we leave implicit the operator nature of spatial variables such as the atomic position $x$.
Each state can be detuned in energy by $\delta_j$ from Raman resonance, giving the contribution to the Hamiltonian 
\begin{equation}
\hat{U}=\sum_{j=0}^{N-1}\delta_{j}|j\rangle\langle j|. \label{eq:U-definition}
\end{equation}
Finally including the kinetic energy yields the full Hamiltonian
\begin{align}
\hat{H} &=\frac{p^{2}}{2m}+\hat{V}\left(x\right)+\hat{U}\,, \label{eq:H-original}
\end{align}
where $p = - \mathrm{i} \partial_x$ is the momentum operator, and in what follows we take $\hbar=1$.

\subsection{Dressed state basis}

Because the internal states $\ket{j}$ can be interpreted as sites in a synthetic dimension~\cite{Boada2012,Celi2014}, it is convenient to adopt a synthetic ``momentum'' representation, giving a new basis of (position independent) dressed states 
\begin{equation}
|\varepsilon_{n}\rangle=\frac{1}{\sqrt{N}}\sum_{j=0}^{N-1}\ket{j} e^{\mathrm{i}2\pi nj/N}\,,\quad\mathrm{with}\quad n=0,1,\ldots,N-1\,.\label{eq:varspesilon_n(x)}
\end{equation}
As above we periodically label states implying
$|\varepsilon_{n+N}\rangle=|\varepsilon_{n}\rangle$.

The light-matter coupling operator [Eq.~\eqref{eq:V-original}] can be represented in the basis of dressed states as
\begin{equation}
\hat{V}\left(x\right)=\sum_{l}\hat{V}_{l}\left(x\right)\,,\label{eq:V-dressed-basis-000}
\end{equation}
with terms
\begin{equation}
\hat{V}_{l}\left(x\right)=-\tilde{\Omega}_{l}\sum_{n=0}^{N-1}{\rm e}^{{\rm i}\left[2\pi\left(n+l\right)/N-qx\right]}|\varepsilon_{n}\rangle\langle\varepsilon_{n+l}|+{\rm H.\,c.}\,\label{eq:V_l}
\end{equation}
resulting from the $l$-th Fourier component of the transition amplitudes
\begin{equation}
\tilde{\Omega}_{l}=\frac{1}{N}\sum_{j=0}^{N-1}\Omega_{j}e^{\mathrm{i} 2\pi l j / N}\,.\label{eq:Omega-tilde_l}
\end{equation} 

\subsection{Dressed state potential $V_{0}\left(x\right)$}

We now consider the situation where the $l=0$ Fourier component is dominant, so
\begin{equation}
\tilde{\Omega}_{0}\gg\tilde{\Omega}_{l}\quad\mathrm{with}\quad l\ne0\,.\label{eq:Omega}
\end{equation}
This component 
\begin{equation}
\Omega\equiv\tilde{\Omega}_{0}=\frac{1}{N}\sum_{j=0}^{N-1}\Omega_{j}
\end{equation}
is the average of the Rabi frequencies $\Omega_{j}$.
The corresponding contribution to $\hat{V}\left(x\right)$ is diagonal in the basis of dressed states
$|\varepsilon_{n}\rangle$ basis, giving
\begin{equation}
\hat{V}_{0}\left(x\right)=\sum_{n=0}^{N-1}\varepsilon_{n}\left(x\right) |\varepsilon_{n}\rangle\langle\varepsilon_{n}|\,,\label{eq:V-dressed-basis}
\end{equation}
where each
\begin{equation}
\varepsilon_{n}\left(x\right)=-2\Omega\cos\left(2 \kr x-2\pi n/N\right)\,\label{eq:varepsilon_n}
\end{equation}
is a sinusoidal potential for atoms moving in $|\varepsilon_{n}\rangle$.
The potentials $\varepsilon_{n\pm1}\left(x\right)$ for the neighboring dressed states $|\varepsilon_{n+1}\rangle$ and $|\varepsilon_{n-1}\rangle$ are each spatially shifted from $\varepsilon_{n}\left(x\right)$ by a distance $a=a_{0}/N$, giving a new unit cell that is $N$ times smaller than the $a_{0}=\lambda/2$ period of a conventional optical lattice.
The dashed curves in Fig.~\ref{fig:setup}(b) illustrate the lattice potentials $\varepsilon_{n}\left(x\right)$ for the case of three internal states ($N=3$).

\subsection{Coupling between dressed states}

\subsubsection{Coupling between dressed states via laser coupling}

The Fourier components $\tilde{\Omega}_{l}$ with $l\ne0$ induce tunable couplings $V_{l}\left(x\right)$ [Eq.~(\ref{eq:V_l})] between atoms in dressed states $|\varepsilon_{n+l}\rangle$ and $|\varepsilon_{n}\rangle$; the corresponding potential minima are separated by a distance $l/N$.
The total contribution of these components is
\begin{equation}
\hat{V}^{\prime}\left(x\right)=\sum_{l\ne0}\hat{V}_{l}\left(x\right)\,.\label{eq:V'(x)}
\end{equation}
Since each $\tilde{\Omega}_{l}$ is a discrete Fourier transform of the coupling matrix element $\Omega_{j}$, changing its $j$-dependence can generate a range of tunneling amplitudes $\tilde{\Omega}_{l}$ that can vary from short to long ranged.
In the following we consider a uniform atom-light coupling, $\Omega_j=\Omega$ and thus $\hat{V}\left(x\right)=\hat{V}_0\left(x\right)$, and concentrate on the effects of the detunings to be considered next. 

\subsubsection{Coupling between dressed states via detuning}

The dressed states are also coupled via inhomogeneous ($j$-dependent) detunings $\delta_{j}$.
In the dressed state basis the detuning operator (\ref{eq:U-definition})
is 
\begin{equation}
\hat{U}=\sum_{n}^{N-1}\sum_{l}U_{l}|\varepsilon_{n}\rangle\langle\varepsilon_{n+l}|\,,\label{eq:U-dressed-basis-1}
\end{equation}
where
\begin{equation}
U_{l}=\frac{1}{N}\sum_{j=0}^{N-1}\delta_{j}\exp\left(\mathrm{i}\frac{2\pi lj}{N}\right)\,\label{eq:U_l}
\end{equation}
describes coupling between dressed states $|\varepsilon_{n}\rangle$ and $|\varepsilon_{n+l}\rangle$ separated by $l$.
The $l=0$ term provides a uniform energy offset and will be omitted. 

Similar to the case of inhomogeneous Rabi frequencies,  the coupling matrix element $U_{l}$ between dressed states $|\varepsilon_{n}\rangle$ and $|\varepsilon_{n+l}\rangle$ is a discrete Fourier transform of the detunings $\delta_{j}$.
Therefore $U_{l}$ can achieve a desired long-range structure on demand by properly choosing the $j$-dependence of $\delta_{j}$.  

It is useful to represent Eq.~\eqref{eq:H-original} for the full Hamiltonian as 
\begin{align}
\hat{H}&=\hat{H}_{0}+\hat{U}\,,\label{eq:H-separation-1}
\end{align}
where the zero order Hamiltonian
\begin{equation}
\hat{H}_{0}=\frac{p^2}{2m}+\hat{V}_{0}\left(x\right)\,\label{eq:H_0}
\end{equation}
consists of the kinetic energy operator and the dressed state potential $\hat{V}_{0}\left(x\right)$ defined by Eq.~\eqref{eq:V-dressed-basis}.
In what follows we treat the detuning operator $\hat{U}$ as a perturbation which couples
the dressed states. 

\subsection{Coupling dressed states with sinusoidal detuning}

When the detuning 
\begin{equation}
\delta_{j,l}=2\delta\cos\left(2\pi l j/N-\varphi\right)\,,\label{eq:delta_j-oscil-detun}
\end{equation}
is a sinusoidal function of the internal state index $j$, the detuning operator \eqref{eq:U-dressed-basis-1} takes the simplified form
\begin{equation}
\hat{U}_l=\delta\sum_{n=0}^{N-1}|\varepsilon_{n}\rangle\langle\varepsilon_{n+l}|e^{\mathrm{i}\varphi}+H.c.\,,\label{eq:U-dressed-basis-oscil-detun-alternative}
\end{equation}
that couples dressed states $|\varepsilon_{n}\rangle$ and $|\varepsilon_{n+l}\rangle$ separated by $l$ ``sites'' in a synthetic dimension picture.

In what follows we consider detunings of the form
\begin{equation}
\delta_{j}=2\sum_{p}\delta^{(p)}\cos\left(2\pi j/N-\varphi^{(p)}\right)\,,\label{eq:delta_j-oscil-detun-1}
\end{equation}
coupling only neighboring dressed states (i.e. $l=1$, so we suppress the $l$ index) with time dependent phases $\varphi^{(p)}\equiv \varphi^{(p)}(t)$ and amplitudes $\delta^{(p)}$. 
In the remainder of this manuscript we focus on the specific case of three phases 
\begin{equation}
\varphi^{(p)}(t) = p\omega t+\gamma^{(p)}\label{eq:varphi^p}\quad (p=0,\pm1)\,,
\end{equation}
with drive frequency $\omega$  and phase shifts $\gamma^{(p)}$.
This leads to the detuning operator
\begin{equation}
\hat{U}=\sum_{n=0}^{N-1} F(t)|\varepsilon_{n}\rangle\langle\varepsilon_{n+1}|+ {\rm H.c.}\,,\label{eq:U-dressed-basis-oscil-detun-many-deltas}
\end{equation}
where
\begin{equation}
F\left(t\right)=\sum_{p=0,\pm1}\delta^{(p)} e^{\mathrm{i}p\omega t+\mathrm{i}\gamma^{(p)}}\,\label{eq:F-three-deltas}
\end{equation}
contains a time-independent component ($p=0$), and a pair of components ($p=\pm1$) with oscillatory exponents $\propto\exp\left(\pm \mathrm{i} \omega t\right)$.

\section{Symmetries and Bloch states}\label{sec:Symmetries}

\subsection{Spatial shift by $a_{0}$ and Bloch solutions}

The complete state vector of the system is
\begin{equation}
\ket{\psi} \equiv \int dx \ket{\psi(x)}\otimes\ket{x}
\label{state-vector}
\end{equation}
where $\ket{\psi(x)} = \braket{x}{\psi}$ is the state vector of the atomic internal states at the position $x$, and $\ket{x}$ is eigenvector of the position operator. 

The atom-light interaction operator $\hat{V}\left(x\right)=\hat{V}\left(x+a_{0}\right)$
in Eq.~(\ref{eq:V-original}) has an obvious spatial periodicity of $a_{0}=2\pi/k$.
As a result the Hamiltonian $\hat{H}$ commutes with the spatial displacement
operator $T(\xi)\equiv \exp(i p_x \xi)$, for $\xi = a_0$, i.e., $[\hat{H}, T(a_0)]=0$.
The operators $\hat{H}$ and $T(a_{0})$ can be simultaneously diagonalized, giving Bloch states 
\begin{align}
\ket{\psi^{(k)}(x)} &= e^{\mathrm{i}kx} \ket{g^{(k)}(x)} \,,\label{eq:psi-q}
\end{align}
where
\begin{align}
\ket{g^{(k)}(x+a_{0})} &= \ket{g^{(k)}(x)} \,\label{eq:g^q-periodic part}
\end{align}  
is the spatially periodic contribution and the crystal-momentum $k$ lies within the standard Brillouin zone (BZ) $-\pi/a_{0}\leq k<\pi/a_{0}$. 
In the next section we identify an additional symmetry that reduces the unit cell to $a = a_0 / N$.

\subsection{Spatial shift by $a$}

In addition to being invariant with respect to a spatial translation of $a_0$, the Hamiltonian $\hat{H}$ commutes with a combined translation operator
\begin{align}
\hat{S}(\xi) &= \hat{Q} T(\xi),\label{eq:T_a}
\end{align}
when $\xi = a$.
The operator $T(a)$ implements a spatial shift by a distance $N$ times smaller than the original lattice constant $a_{0}$. 

The shift is accompanied by a change in the atomic internal state described by 
\begin{equation}
\hat{Q}=\sum_{j=0}^{N-1}\left|j\right\rangle \left\langle j\right|{\rm e}^{{\rm i}j2\pi/N}=\sum_{n=0}^{N-1}\left|\varepsilon_{n+1}\right\rangle \left\langle \varepsilon_{n}\right|\,.\label{eq:Q}
\end{equation}
This is the synthetic dimension displacement operator for dressed state basis.
The combined symmetry operator $\hat{S}(a)$ is closely related to the magnetic displacement operators that appear in the study of the Hofstadter model of charged particles in a square lattice~\cite{Zak1964,Hofstadter1976,Bernevig2013}.

Because the Hamiltonian $\hat{H}$ is invariant with respect to the combined shift $\hat{S} (a)$, the operators $\hat H$ and $\hat{S} (a)$ have a common set of eigenstates analogous to the Bloch states in Eq.~\eqref{eq:psi-q}, but with $N$ times smaller periodicity $a = a_0 / N$.
Thus the eigenvectors $\ket{\psi^{(k)}(x)}$
of $\hat{S}(a)$ have eigenvalues $\exp(i k a)$ characterised by a crystal-momentum $k$ covering an $N$-fold enlarged BZ with $-N\pi/a_{0} \leq k< N\pi/a_{0}$.

The periodic part of these states have the property
\begin{align}
\hat{Q}\ket{g^{(k)}(x+a)} &= \ket{g^{(k)}(x)}\,. \,\label{eq:Q-g^q-eigenvalue}
\end{align}
so that
\begin{align}
\ket{g^{(k)}(x)} &= \sum_{n=1}^{N} g^{(k)}(x+na)\ket{\varepsilon_{n}}\,,\label{eq:g^q_n-expanded-dressed}
\end{align}
with $g^{(k)}(x+Na)=g^{(k)}(x)$.

\section{Tight binding approach}\label{sec:TightBinding}

\subsection{Wannier functions}

\begin{figure}[tbh]
\includegraphics{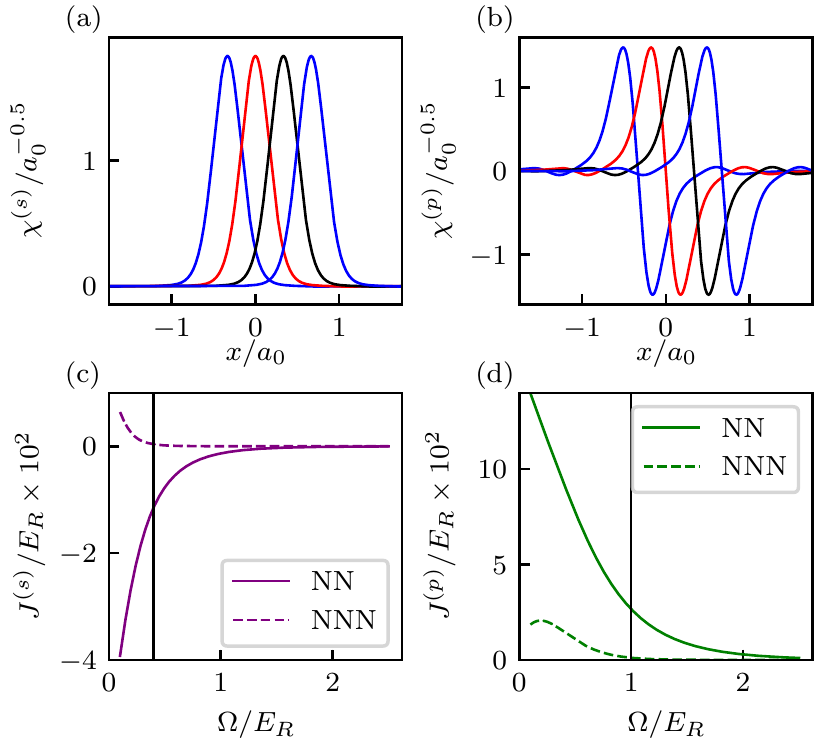}
\caption{Two band model.
(a) and (b) Wannier functions for the $s$-band and $p$-band respectively, computed for $\Omega=1\Er$, $\delta=0\Er$, and $N=3$.
The colors correspond to Wannier states for each of the three dressed states. 
(c) and (d) Natural tunneling $J^{(\alpha)}$ dependence on Rabi frequency $\Omega$ for the $s$- and $p$- bands. 
The solid and dashed curves plot the nearest neighbor (NN) and next nearest neighbor tunneling (NNN). 
The black lines denote the threshold for the applicability of the tight binding approximation: for $\alpha=s$ this threshold is $\Omega>0.4\Er$, and for $\alpha=p$ it is $\Omega>1.0\Er$. 
}
\label{fig:combo}
\end{figure}

We begin by considering maximally localized Wannier states associated with each dressed state potential $\varepsilon_{n}\left(x\right)$ in the zero-order
Hamiltonian $\hat{H}_{0}$ given by Eq.~\eqref{eq:H_0}. 
The Wannier functions $\chi^{(\alpha)}\left(x-ar\right)$ for the $n$-th dressed state are localized around each local minima of $\varepsilon_{n}\left(x\right)$ at $x/a=r=n+Nl$.
The integer $l$ defines a periodic array of lattice sites, and $\alpha=0,1,\ldots$ (or equivalently $\alpha={\rm s},{\rm p},\ldots$ ) labels the different Bloch bands.

The abstract state vector 
\begin{align}
\ket{r,\,\alpha} &= \int dx \chi^{(\alpha)}\left(x-ar\right)
\ket{\varepsilon_{r}} \otimes\ \ket{x}
\,.\label{eq:Wannier-state-vectors}
\end{align}
corresponding to the Wannier function $\chi^{(\alpha)}\left(x-ar\right)$  includes the dressed state $\ket{\varepsilon_{r}}=\ket{\varepsilon_{n+Nl}}=\ket{\varepsilon_{n}}$.
Figure~\ref{fig:combo}(a,b) plots Wannier functions for the lowest two bands and the colors denote the different dressed states.

In the Wannier basis, the combined shift operator becomes
\begin{align}
\hat{S}(a) &= \sum_{r,\,\alpha}\ket{r+1,\,\alpha} \bra{ r,\,\alpha}\,,\label{eq:T_a-Wannier}
\end{align}
and the zero order Hamiltonian reduces to
\begin{align*}
\hat{H}_{0} &=\sum_{\alpha,r}\Bigl[\epsilon^{(\alpha)}\ket{r,\,\alpha}\bra{ r,\,\alpha}\\
&\ \ \ \ + J^{(\alpha)}\Bigl(\ket{r,\,\alpha}\bra{r+N,\,\alpha}+{\rm H.c}.\Bigr)\Bigr]\,,
\end{align*}
in the tight binding limit with only nearest neighbor (NN) tunneling.
Here $J^{(\alpha)}$ is a matrix element for tunneling between NN Wannier functions in the same internal state, and $\epsilon^{(\alpha)}$ is the onsite energy.  
The tight binding approximation holds for the lowest two bands (the $s-$ and $p-$bands corresponding to $\alpha=0$ and $1$) when $\Omega\gtrsim \Er$.
The vanishing of long-range tunneling in these bands is plotted in Fig.~\ref{fig:combo} (c) and (d).

Although each dressed state is subject to a lattice with period $a_0$, the spacing between neighboring Wannier functions is $a=a_{0}/N$.
These adjacent functions are in different dressed states with $\ket{\varepsilon_{n}}$ and $\ket{\varepsilon_{n\pm1}}$.
This spacing is $N$ times smaller than the original lattice constant $a_{0}$.
This provides a one-dimensional lattice [shown by the dashed curves in Fig.~\ref{fig:setup}(b) for $N=3$] with a periodicity $a=a_{0}/N$, in which the ``natural tunneling'' occurs between Wannier functions of the same dressed state separated by $a_0$. 

As we will see Sec.~\ref{subsec:Coupling-between-different}, including the detuning operator $\hat{U}$ as a perturbation introduces tunneling between Wannier functions of different dressed states.
Furthermore as discussed in Sec.~\ref{sec:TimeDependent}, different Bloch bands---such the $s$- and $p$-bands---can be coupled by making the detuning oscillatory in time.  

\subsection{Coupling between Wannier functions via detuning\label{subsec:Coupling-between-different}}

The interaction operator $\hat{U}$ in Eq.~\eqref{eq:U-dressed-basis-oscil-detun-many-deltas} describes detuning induced coupling between neighboring atomic internal dressed states $\ket{\varepsilon_{n}}$ and $\ket{\varepsilon_{n+1}}$. 
In the Wannier basis [defined in Eq.~(\ref{eq:Wannier-state-vectors})], the leading order contribution to $\hat{U}$ is
\begin{equation}
\hat{U}=F\left(t\right)\sum_{r, \alpha,\alpha^{\prime}}|r,\,\alpha\rangle G_{\alpha,\alpha^{\prime}}\langle r+1,\,\alpha^{\prime}|+{\rm H.c.}\,,\label{eq:U-dressed-basis-Wannier-NN}
\end{equation}
where
\begin{equation}
G_{\alpha,\alpha^{\prime}}=\int_{-\infty}^{+\infty}\chi^{(\alpha)*}\left(x-a\right)\chi^{(\alpha^{\prime})}\left(x\right)\mathrm{d}x\,\label{eq:G-NN}
\end{equation}
is the overlap integral between the neighboring Wannier functions.

When the detuning is small compared to the energy difference between Bloch bands, $\hat{U}$ can be
treated as a perturbation that induces transitions between spatially separated Wannier functions in different dressed states. 
For deep lattices (e.g. when $\Omega\gg\Er$), the matrix element of the direct tunneling $J^{(\alpha)}$ can be negligible in the lowest Bloch bands.
This allows detuning-induced coupling between Wannier functions spaced by $a$ (i.e., coupling neighboring dressed states) to become the dominant source of tunneling.

\section{Interband coupling via time-dependent detuning\label{sec:TimeDependent}}
 
The function $F(t)$ entering the detuning 
operator $\hat{U}(t)$ in Eqs.~\eqref{eq:U-dressed-basis-oscil-detun-many-deltas} and \eqref{eq:U-dressed-basis-Wannier-NN} contains a constant term $ \delta^{(0)}$ and two Fourier components $\delta^{(\pm 1)}\exp\left(\pm \mathrm{i} \omega t\right)$.
As we found in the last section, a constant detuning  $ \delta^{(0)}$ generates tunneling between neighboring Wannier functions in the same Bloch band, but for different dressed states. 
As we will see, the oscillatory terms
$\propto \delta^{(\pm 1)}\exp\left(\pm \mathrm{i} \omega t\right)$
can resonantly couple Wannier functions in different Bloch bands and different dressed states $\ket{\varepsilon_{n}}$ and $\ket{\varepsilon_{n+1}}$.

The three-term form of $F(t)$ is in contrast with conventional periodic driving~\cite{Eckardt2017} where the tunneling matrix elements acquire phases such as $\propto \cos\left(\omega t\right)$. 
In that case, the tunneling elements are described by an infinite sum of Fourier components with amplitudes given by Bessel functions; this creates additional possibilities for unwanted coupling to higher Bloch bands.

\subsection{Transition to the rotating frame}

We focus on resonant driving where the energy difference between the ground ($s$) and first excited ($p$) bands is close to the driving frequency, i.e., 
\begin{align*}
\left|\epsilon^{(p)}-\epsilon^{(s)}-\omega\right| &\ll \omega.
\end{align*}
In this limit it is convenient to transform the tight binding Hamiltonian to the rotating frame with the unitary transformation
\begin{align}
\hat{S} &= \exp\left(-\mathrm{i} \omega t \sum_{\alpha,r}\alpha \ket{r,\,\alpha}\bra{r,\,\alpha} \right)\,.\label{eq:S}
\end{align}
The transformed Hamiltonian is 
\begin{equation}
\tilde{H}=\tilde{H}_{0}+\tilde{U}\,;\label{eq:H-tilde-full}
\end{equation}
because $\hat{S}$ commutes with the zero order Hamiltonian $\hat{H_0}$ but not $\tilde{U}$ we have
\begin{equation}
\tilde{H}_{0}=\hat{H}_0-\mathrm{i}\hat{S}^{\dagger}\partial_{t}\hat{S}\,\quad\mathrm{and}\quad\tilde{U}=\hat{S}^{\dagger}\hat{U}\hat{S}\,,\label{eq:H_0-tilde,U-tilde}
\end{equation}
where tildas mark transformed operators.
In the Wannier basis the transformed operators are

\begin{align*}
\tilde{H}_{0} &= \sum_{\alpha,r}\Bigl[\tilde{\epsilon}^{(\alpha)}\ket{r,\,\alpha}\bra{ r,\,\alpha}\\
&\ \ \ \ +J^{(\alpha)}\Bigl(\ket{r,\,\alpha}\bra{ r+N,\,\alpha}+{\rm H.c.}\Bigr)\Bigr]\,,
\end{align*}
and 
\begin{align*}
\tilde{U} &= \sum_{\alpha,\alpha^{\prime},r}\Bigl[G_{\alpha,\alpha^{\prime}} F(t)e^{\mathrm{i}(\alpha^{\prime}-\alpha)\omega t}\ket{r,\,\alpha} \bra{ r+1,\,\alpha^{\prime}}\\
&\ \ \ \ \ + {\rm H.c.}\Bigr]\,.
\end{align*}
where
\begin{equation}
\tilde{\epsilon}^{(\alpha)}=\epsilon^{(\alpha)}-\alpha\omega\label{eq:detuning}
\end{equation}
are the shifted energies of the Bloch bands. 

\subsection{Rotating wave approximation}

We now make the rotating wave approximation (RWA) by omitting the oscillating terms remaining in the transformed operator $\tilde{U}$.
We express the RWA detuning operator as a sum
\begin{equation}
\tilde{U}=\tilde{U}_{0}+\tilde{U}_{1}+\tilde{U}_{-1}\,,\label{eq:U-tilde-separation}
\end{equation}
over three terms
\begin{align}
\tilde{U}_{0} &= \sum_{\alpha,r}\left[G_{\alpha,\alpha}\delta^{(0)}e^{\mathrm{i}\gamma^{(0)}} \ket{r,\,\alpha}\bra{r+1,\,\alpha} + {\rm H.c.}\right] \label{eq:U-tilde} \\
\tilde{U}_{1} &= \sum_{\alpha,r} \Bigl[G_{\alpha,\alpha-1}\delta^{(1)}e^{\mathrm{i}\gamma^{(1)}} \ket{r,\,\alpha}\bra{r+1,\,\alpha-1}\\
&\ \ \ \ + {\rm H.c.}\Bigr] \nonumber \\
\tilde{U}_{-1} &= \sum_{\alpha,r}\Bigl[G_{\alpha,\alpha+1}\delta^{(-1)}e^{\mathrm{i}\gamma^{(-1)}} \ket{r,\,\alpha}\bra{r+1,\,\alpha+1}\\
&\ \ \ \ + {\rm H.c.}\Bigr]\,.\nonumber
\end{align}
where each term results from the corresponding term in $F\left(t\right)$.
The time independent detuning $\delta^{(0)}$ leads to $\tilde{U}_{0}$, which as discussed above, describes tunneling between neighboring Wannier functions in the same Bloch band $\alpha$.
On the other hand, $\tilde{U}_{\pm1}$ describe tunneling between
neighboring Wannier functions in consecutive Bloch bands, where a transition from $\alpha$ to $\alpha\pm1$ is accompanied by moving from site $r$ to $r-1$ (and vice versa). 
These two processes are independent, as they are separately controlled the amplitudes of the oscillating detunings $\delta^{(1)}$ and $\delta^{(-1)}$. 
Additionally, changing the drive frequency $\omega$ alters the RWA energy offsets of the Bloch bands $\tilde{\epsilon}^{(\alpha)}$.

\begin{figure}[tb!]
\includegraphics[width=1.00\columnwidth]{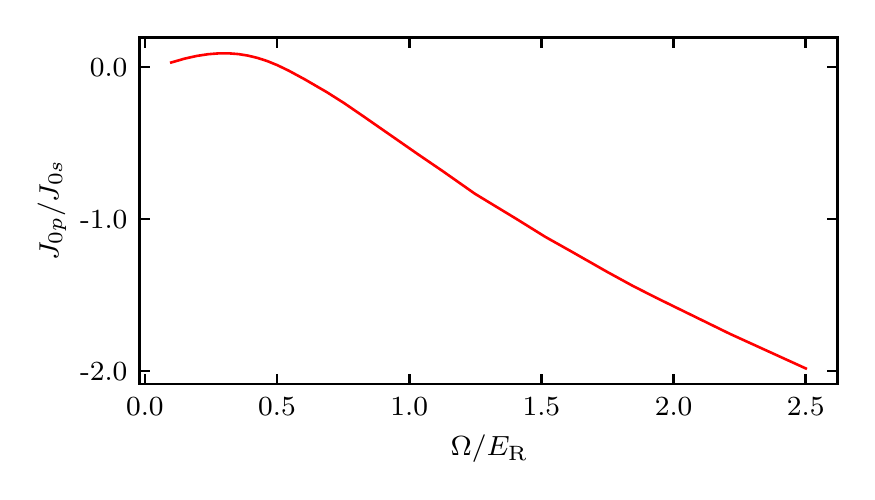}
\caption{Modulation induced tunneling ratio
$J_{0p}/J_{0s}$ calculated exactly, plotted as a function of $\Omega$.}
\label{fig:t_0p--t_0s_ratio}
\end{figure}

We now specialize to the case where only the lowest two bands (identified by $\alpha = s$ or $p$) are coupled, in which case the components of $\tilde U$ reduce to
\begin{align*}
\tilde{U}_{0} &= \sum_{\alpha=s,p}\sum_{r} J_{0,\alpha}e^{\mathrm{i}\gamma^{(0)}} \ket{r,\,\alpha}\bra{ r+1,\,\alpha} + {\rm H.c.} \\
\tilde{U}_{1} &= \sum_{r}  J_{1}e^{\mathrm{i}\gamma^{(1)}} \ket{r,\,p}\bra{r+1,\,s} + {\rm H.c.} \\
\tilde{U}_{-1} &= \sum_{r} J_{-1}e^{\mathrm{i}\gamma^{(-1)}} \ket{r,\,s}\bra{r+1,\,p} + {\rm H.c.}\,,
\end{align*}
with constants
\begin{align*}
J_{0\alpha}\!&=\!G_{\alpha,\alpha}\delta^{(0)}, & J_{\pm1}\!&=\!\pm G_{p,s}\delta^{(\pm1)}, & {\rm and} &&  G_{p,s} \!&=\!-G_{s,p}\,.
\end{align*}
We find that $G_{s,s}$ is strictly positive, while $G_{p,p}$ is negative for $\Omega\gtrsim 1\Er$, as indicated by the ratio $J_{0p}/J_{0s} = G_{p,p}/G_{s,s}$ plotted in Fig.~\ref{fig:t_0p--t_0s_ratio}.
We also note that for $\Omega\lesssim 1\Er$ coupling to higher bands cannot be neglected.

All together, this realizes a pair of RM chains [vertical lines in Fig.~\ref{fig:setup}(c)] coupled by the static detuning $\delta_{0}$ [diagonal lines in Fig.~\ref{fig:setup}(c)].
The resulting physics, going beyond that of the RM model, is the focus of the remainder of this manuscript. 

\subsection{Coupled Rice-Mele chains}

Because the full RWA Hamiltonian commutes with the combined translation operator $\hat S (a)$ given by Eq.~\eqref{eq:T_a-Wannier}, the eigenstates $\ket{k,\beta}$ are labeled by crystal momentum $k$ (covering the extended Brillouin zone $-2\pi/a\le k<2\pi/a$) as well as a band index $\beta$.
Because the periodic modulation couples the initial bands (labeled by $\alpha$), the eigenstates take the form
\begin{align}
\ket{k,\beta} &= \sum_{\,\alpha}c^{(\beta)}_{\alpha,\,k}\ket{k,\,\alpha}\,,\label{eq:|k>-Wannier_basis-1}
\end{align}
in terms of the eigenstates
\begin{align*}
\ket{k,\,\alpha} &= \frac{1}{L^{1/2}}\sum_{r}\ket{r,\,\alpha} e^{\mathrm{i}kra}\,
\end{align*}
of $\tilde H_0$ for a system $L$ sites in extent. 

When we consider only the lowest two bands, i.e. $\alpha \in \left\{s,\,p\right\}$, Eq.~\eqref{eq:|k>-Wannier_basis-1} reduces to
\begin{align}
\ket{k, \beta} &= c^{(\beta)}_{s,\,k}\ket{k,\,s} +c^{(\beta)}_{p,\,k}\ket{k,\,p} \,.\label{eq:|k>-s,p}
\end{align}
with $\beta = \pm$.
Therefore the eigenvalue equation $\tilde{H}\ket{k,\beta} = E_{k,\beta}\ket{k,\beta}$
can be expressed as a $2\times2$ matrix equation
\begin{equation}
\tilde H_{k}\begin{pmatrix}c^{(\beta)}_{s,\,k}\\
c^{(\beta)}_{p,\,k}
\end{pmatrix} = E_{k,\beta}\begin{pmatrix}c^{(\beta)}_{s,\,k}\\
c^{(\beta)}_{p,\,k}
\end{pmatrix}\,,\label{eq:2x2matrix-eigen-equation}
\end{equation}
where $\tilde H_{k}$ is the Hamiltonian matrix
\begin{equation}
\tilde H_{k}=\begin{pmatrix}\Lambda_{s,k} & \Omega_{k}^{*}/2\\
\Omega_{k}/2 & \Lambda_{p,k}
\end{pmatrix}\,.\label{eq:H_k}
\end{equation}
The off diagonal matrix elements 
\begin{align*}
\frac{\Omega_{k}}{2} &= J_{1}\exp\left[\textrm{i}\left(ka+\gamma^{(1)}\right)\right] \\
&\ \ \ +J_{-1}\exp\left[-\textrm{i}\left(ka+\gamma^{(-1)}\right)\right]. 
\end{align*}
that couple the bands are due to the modulated detuning $\delta_{\pm 1}$.
On the other hand, the diagonal matrix elements
\begin{align*}
\Lambda_{\alpha,k} &= \tilde{\epsilon}^{(\alpha)} + 2J^{(\alpha)}\cos\left(kNa\right) + 2J_{0,\alpha}\cos\left(ka+\gamma^{(0)}\right)\,, 
\end{align*}
are due to the static detuning $\delta_{0}$ and the natural tunneling $J^{(\alpha)}$.

In what follows we fix the modulation phases to be $\gamma^{(\pm1)}=0$ and $\gamma^{(0)}=-\pi/2$ and define the energy shift and detuning
\begin{align}
\Lambda_{k} &= \frac{\Lambda_{s,k}+\Lambda_{p,k}}{2} & \mathrm{and} && \Delta_{k} &= \Lambda_{s,k}-\Lambda_{p,k}\,.\label{eq:Delta_k}
\end{align}
Subtracting the overall $\Lambda_{k}$ energy shift gives
\begin{equation}
\tilde H_{k}=\frac{1}{2}\begin{pmatrix}\Delta_{k} & \Omega_{k}^{*}\\
\Omega_{k} & -\Delta_{k}
\end{pmatrix}\,.\label{eq:h_k}
\end{equation}
Here $\Delta_{k}$ and $\Omega_{k}$ are
\begin{equation}
\Delta_{k}=\tilde{\epsilon}+2J \cos\left(kNa\right)+2J_{0}\sin\left(ka\right)\,.\label{eq:Delta_k-specific}
\end{equation}
and
\begin{equation}
\frac{\Omega_{k}}{2}=J_{1}e^{\textrm{i}ka}+J_{-1}e^{-\textrm{i}ka}\,\label{eq:Omega_1k-specific}
\end{equation}
with
\begin{align*}
\tilde{\epsilon} &= \tilde{\epsilon}^{(s)} - \tilde{\epsilon}^{(p)}, & J & = J^{(s)}-J^{(p)}, & {\rm and} && J_{0} &= J_{0,s}-J_{0,p}\,.
\end{align*}
For most parameters $J_{0,s}$ and $J_{0,p} < 0$, so in what follows we take $J_0 \geq 0$.

The eigenenergies of $\tilde H_{k}$ are
\begin{align}
E_{k,\pm} &= \pm\sqrt{\Delta_{k}^{2}+\left|\Omega_{k}\right|^{2}}\,,\label{eq:Omega_k}
\end{align}
and the corresponding eigenstates~\cite{Dalibard2011,Xiao2010}
\begin{align*}
|k,-\rangle &= \left(\begin{array}{c}
e^{-\mathrm{i}\phi}\sin\left(\theta/2\right)\\
-\cos\left(\theta/2\right)
\end{array}\right), & |k,+\rangle &= \left(\begin{array}{c}
e^{-\mathrm{i}\phi}\cos\left(\theta/2\right)\\
\sin\left(\theta/2\right)
\end{array}\right)\,,
\end{align*}
can be represented in terms of the angles $\theta$ and $\phi$ given by 
\begin{equation}
e^{\mathrm{i}\phi}=\Omega_{k}/\left|\Omega_{k}\right|\,\quad\mathrm{and}\quad\cos\theta=2\Delta_{k}/\Omega_{k}\,.\label{eq:phi,theta}
\end{equation}

\section{Adiabatic pumping}\label{sec:Pumping}

Here we consider the response of the system when one or more parameters are changed adiabatically and periodically in time with period $T$, making the eigenstates $\ket{k,\pm; t}$ explicit functions of time.
In what follows the time dependence will be implicitly assumed.

\subsection{Berry connection}\label{subsec:Berry}

In general any state $\ket{\psi({\bf u})}$ parameterized by a set of variables $u_j$ can by characterized by the geometric vector potential (Berry connection)
\begin{equation}
A_j = \mathrm{i} \bra{\psi({\bf u})}\partial_{j}\ket{\psi({\bf u})}\,.
\end{equation}
In the present case we consider the eigenstates $\ket{k,\pm}$ of $\tilde H_k$ (which can only be unambiguously defined in the absence of degeneracies), giving 
\begin{equation}
A^{(\pm)}_{j}=\mp\sin^{2}\left(\theta/2\right)\partial_{j}\phi\,,\label{eq:vect}
\end{equation}
with the parameters $u_j = (k,t)$.
Because $\ket{k,\pm}$ is periodic in both $k$ and $t$, ${\bf u}$ describe a torus embedded a 3-dimensional space.
We correspondingly introduce a fictitious coordinate normal to the surface giving
\begin{equation}
\mathbf{A}^{(\pm)}=\left(A^{(\pm)}_{k},A^{(\pm)}_{t},0\right)=\mp\frac{1}{2}\left(1-\cos\theta\right)\boldsymbol{\nabla}\phi\,.\label{eq:A}
\end{equation}
The resulting geometric magnetic field (Berry curvature) is
\begin{equation}
\mathbf{B}^{(\pm)}=\boldsymbol{\nabla}\times\mathbf{A}^{(\pm)}=\left(0,0,B\right)=\mp\frac{1}{2}\boldsymbol{\nabla}\phi\times\boldsymbol{\nabla}\cos\theta\,.\label{eq:B}
\end{equation}
We will focus exclusively on the lower band in the remainder of this document and therefore omit the $\pm$ superscript on ${\bf A}$ and ${\bf B}$ and take the lower sign in Eq.~\eqref{eq:B}.

\subsection{Chern number and Zak phase}

We now consider the properties of this system when Hamiltonian parameters ${\bf u}(t)$ cyclically follow a closed path in parameter space.
It is convenient (although not strictly necessary) to assume that this process is time-periodic with period $T$.
In this case, and for a filled Bloch band, the system undergoes quantized pumping~\cite{Thouless1983,Xiao2010} described by an integer
\begin{equation}
C=\int_{0}^{T}\frac{\mathrm{d}t}{2\pi}\int_{-\pi/a}^{\pi/a}\mathrm{d}kB_{z}
,\label{eq:Chern-definition}
\end{equation}
quantifying the per-cycle spatial displacement in units of the reduced lattice period $a$.

It is convenient evaluate this integral using Stokes' formula to replace the planar integral in Eq.~\eqref{eq:Chern-definition} with a line-integral.
The contribution by the boundary integration  [Fig.~\eqref{fig:integration}] vanishes, because the vector potential $\mathbf{A}$ has periodicity $2\pi/a$ with respect to $k$ and periodicity $T$ with respect to $t$.
Because, the Stokes integration contour must avoid the singular points of $\mathbf{A}$, applying Stokes' formula to Eq.~\eqref{eq:Chern-definition} with Eq.~\eqref{eq:A} for $\mathbf{A}$ yields
\begin{equation}
C=-\frac{1}{4\pi}\sum_{\mathrm{sing}}\oint\left(1-\cos\theta\right)\boldsymbol{\nabla}\phi\cdot\mathrm{d}\boldsymbol{\ell}\,. \label{eq:Chern-singlular-points-explicit}
\end{equation}
The sum runs over the singular
points in $\mathbf{A}$ and the integration around these points is counterclockwise; the overall minus sign results from the clockwise orientation of the original trajectories encircling the singular points; and the differential line element is $\mathrm{d}\boldsymbol{\ell}=\left(\mathrm{d}k,\mathrm{d}t,0\right)$. 

Adiabatic pumping can also be described by the Zak phase~\cite{Zak1989}
\begin{equation}
\gamma_\mathrm{Zak} = \int_{-\pi/a}^{\pi/a}\mathrm{d}k\, A_k\,.
\label{eq:Zak-definition}
\end{equation}
In general $a \gamma_\mathrm{Zak}/(2\pi)$ gives the displacement of each Wannier function's mean position from an initial point (the selection of initial position is arbitrary and its selection behaves like a gauge fixing condition).
Therefore changing the Zak phase by $2\pi$ leads to an overall spatial translation by the reduced lattice constant $a$: quantized pumping.

\subsection{Modulation schemes}\label{Mod schemes}

Here we describe two specific adiabatic modulation schemes leading to topological charge pumping.
Both of these schemes involve modulations of $J_{\pm1}$.
The first, which we call ``the $\epsilon$ scheme'', in addition modulates $\tilde\epsilon$; and the second, called ``the $J_0$ scheme'', (unsurprisingly) instead adds modulation to $J_0$.

We begin by considering the implications of modulating $J_{\pm1}$ for the computation of the Chern number via Eq.~\eqref{eq:Chern-singlular-points-explicit}.
The singular points appearing in this expression occur when
$\Omega_{k}=\left|\Omega_{k}\right|e^{\mathrm{i}\phi} = 0$ and thus are located at $J_{1}=J_{-1}$ and $ka=\pm\pi/2$.

\begin{figure}[tb!]
\includegraphics[width=1.00\columnwidth]{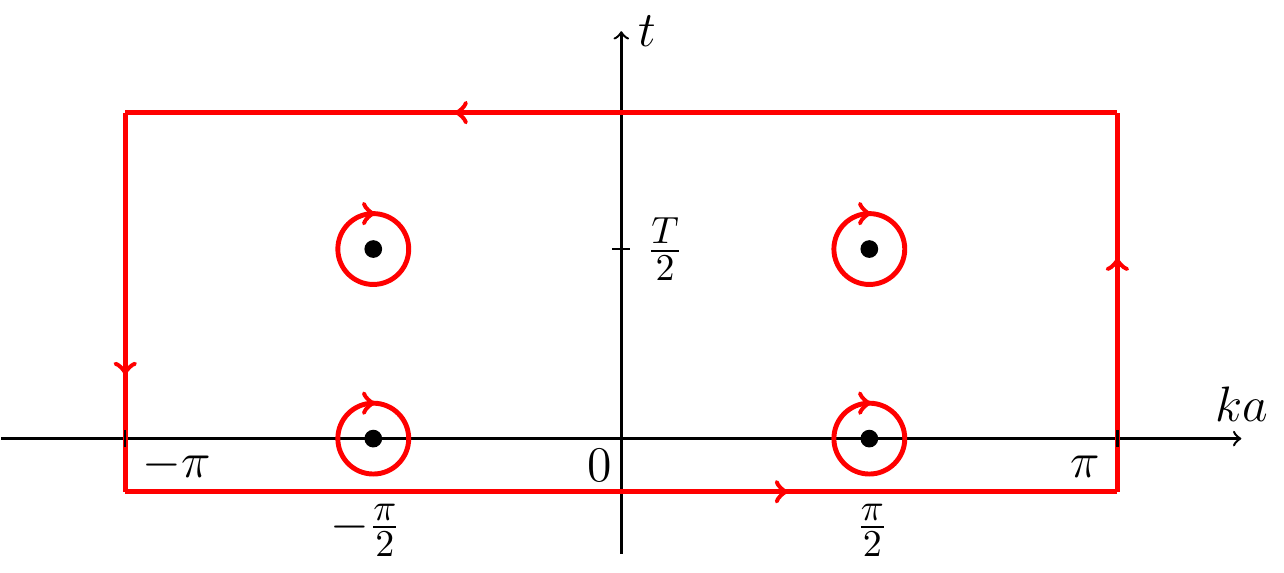}
\caption{Contours of integration. 
Due to the periodicity of the vector potential the boundary integration vanishes.
Thus the integral is fully defined by the small integrals around the excluded singular points in ${\bf A}$ given by Eq.~\eqref{eq:k,t-zero-points}. 
}
\label{fig:integration}
\end{figure}

Let us assume that the $s\rightarrow p$ tunneling elements $J_{1}$ and $J_{-1}$ are modulated with opposite phases
\begin{equation}
J_{\pm1}=\bar{J} \pm J_\mathrm{R}\sin\left(2\pi t/T\right)
\,,\label{eq:t_pm-modulation}
\end{equation}
where $J_\mathrm{R}\ge 0$.
In this case, the only time-dependent reduced parameter in Eq.~\eqref{eq:h_k} is
\begin{align*}
\Omega_{k} &= 4 J\cos (ka)+\mathrm{i}4 J_\mathrm{R} \sin\left(2\pi t/T\right)\sin (ka).
\end{align*}
The function $\Omega_{k}$ is zero at four points shown in Fig.~\ref{fig:integration} corresponding to the four possible combinations of
\begin{equation}
ka=\pm\pi/2
\quad\mathrm{and}
\quad t=jT/2\,,
\quad\mathrm{with}\quad 
j=0,1.\label{eq:k,t-zero-points}
\end{equation}
Integrating the phase gradient around each of these singular points yields
\begin{equation}
C=\frac{1}{2}\sum_{\pm}\sum_{j=0}^{1}\left(-1\right)^{j}\left[\mathrm{sgn}(\Delta_{\pm ,j})-1\right]\,,\label{eq:Chern-singular points}
\end{equation}
via Eq.~\eqref{eq:Chern-singlular-points-explicit}.
Here
where $\mathrm{sgn(}...)$ is the sign function and 
\begin{equation}
\Delta_{\pm,j}=\tilde{\epsilon} 
\pm 2J_{0}\,\label{eq:Delta_pm-explicit}
\end{equation}
is the detuning $\Delta_{k}$ 
given by Eq.~(\ref{eq:Delta_k-specific})
at $ka=\pm\pi/2$ and for $t=jT/2$.

In writing Eq.~\eqref{eq:Delta_pm-explicit} we  omitted the second term in Eq.~\eqref{eq:Delta_k-specific} which is equal to zero for odd values of $N$.
However, for even $N$ it provides a small level shift which can be included into the detuning $\tilde{\epsilon}$. 

\subsubsection{$\epsilon$ modulation scheme}

Here we consider the impact of additionally modulating the detuning
\begin{equation}
\tilde{\epsilon}=
\bar{\epsilon} +
\epsilon_\mathrm{R} \cos\left(2\pi t/T\right)
\,,\label{eq:epsilon_01-modulation}
\end{equation}
around a central value $\bar{\epsilon}$ with extent $\epsilon_\mathrm{R}\ge 0$. 
For $t=jT/2$,
Eq.~(\ref{eq:Delta_pm-explicit}) reduces to
\begin{equation}
\Delta_{\pm ,j}=\bar{\epsilon}+\epsilon_\mathrm{R}\left(-1\right)^{j}\pm 2J_{0}\,.\label{eq:Delta_pm-explicit-1}
\end{equation}

\begin{figure}[tbh!]
\includegraphics[width=0.48\textwidth]{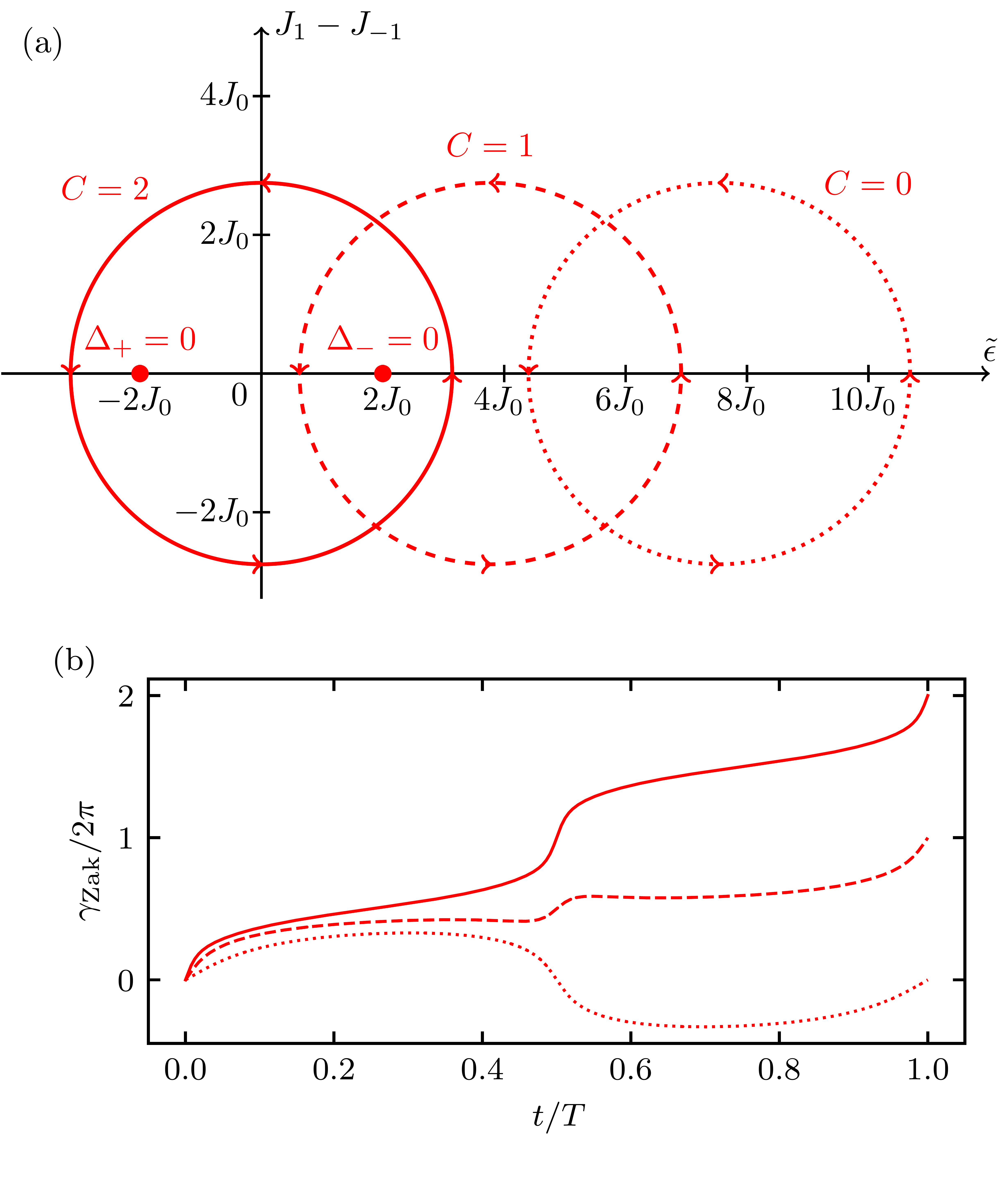}
\caption{Adiabatic pumping in the $\epsilon$ scheme. (a) The three circles show the cases when: both critical points $(\pm 2J_0, 0)$ are encircled (solid),  one of them is encircled (dashed) or neither is encircled (dotted). (b) Zak phase $\gamma_\mathrm{Zak}$ dependence on time $t$ for the three aforementioned trajectories.}
\label{fig:Zak_Combo-epsilon}
\end{figure}

The red points in Fig.~\ref{fig:Zak_Combo-epsilon}(a) mark the singular points in the $\tilde\epsilon-(J_1 - J_{-1})$ plane and the circles show parametric trajectories in this plane.

As shown by the solid circle, both singular points are enclosed when 
\begin{equation}
|\epsilon_\mathrm{R}| > |\bar{\epsilon} \pm 2J_0|\,,
\label{Condition-1}
\end{equation}
In this case, $\Delta_{\pm,j}$ alternates sign for even and odd values of $j$ (because $\mathrm{sgn}(\Delta_{\pm,j})=\left(-1\right)^{j}$), so Eq.~\eqref{eq:Chern-singular points} gives $C=2$.
Thus one arrives at the adiabatic pumping displacement of  $2a$ per pump cycle (as with uncoupled RM chains). 

On the other hand, when only one of the conditions in Eq.~\eqref{Condition-1} holds, just one point $(\pm 2 J_0,
0)$ is encircled by the evolution curve [dashed circle in Fig.~\ref{fig:Zak_Combo-epsilon}(a)] and $C=1$. 
This leads to adiabatic pumping of $a$ per cycle, a scenario which is not possible for uncoupled RM chains. 

Finally, when neither condition holds, both points $(\pm 2J_0, 0)$ are outside the evolution curve [dotted circle in Fig.~\ref{fig:Zak_Combo-epsilon}(a)], yielding the
topologically trivial case with $C=0$.
Figure~\ref{fig:Zak_Combo-epsilon}(b) shows the changes of the
Zak phase for three trajectories of Fig.~\ref{fig:Zak_Combo-epsilon}(a) illustrating that the
adiabatic pumping indeed takes place in units of $2a$, $a$ and $0$ for these three cases.

\subsubsection{$J_0$ modulation scheme}

Lastly we consider modulating
\begin{align}
J_{0} &= \bar{J}_{0}+J_{0,\mathrm{R}} \cos\left(2\pi t/T\right) & {\rm with} && J_{0,\mathrm{R}} > 0,\label{eq:t_0-modulation}
\end{align}
rather than $\tilde{\epsilon}$.
Using Eq.~(\ref{eq:Delta_pm-explicit}), we arrive at the detuning
\begin{equation}
\Delta_{\pm,j}=\tilde{\epsilon}\pm2\left[\bar{J}_{0}+J_{0,\mathrm{R}}\left(-1\right)^{j}\right]\,,\label{eq:Delta_pm-specific-3}
\end{equation}
where $\tilde{\epsilon}$ is taken to be constant and without loss of generality, positive.

\begin{figure}[tbh!]
\includegraphics[width=0.48\textwidth]{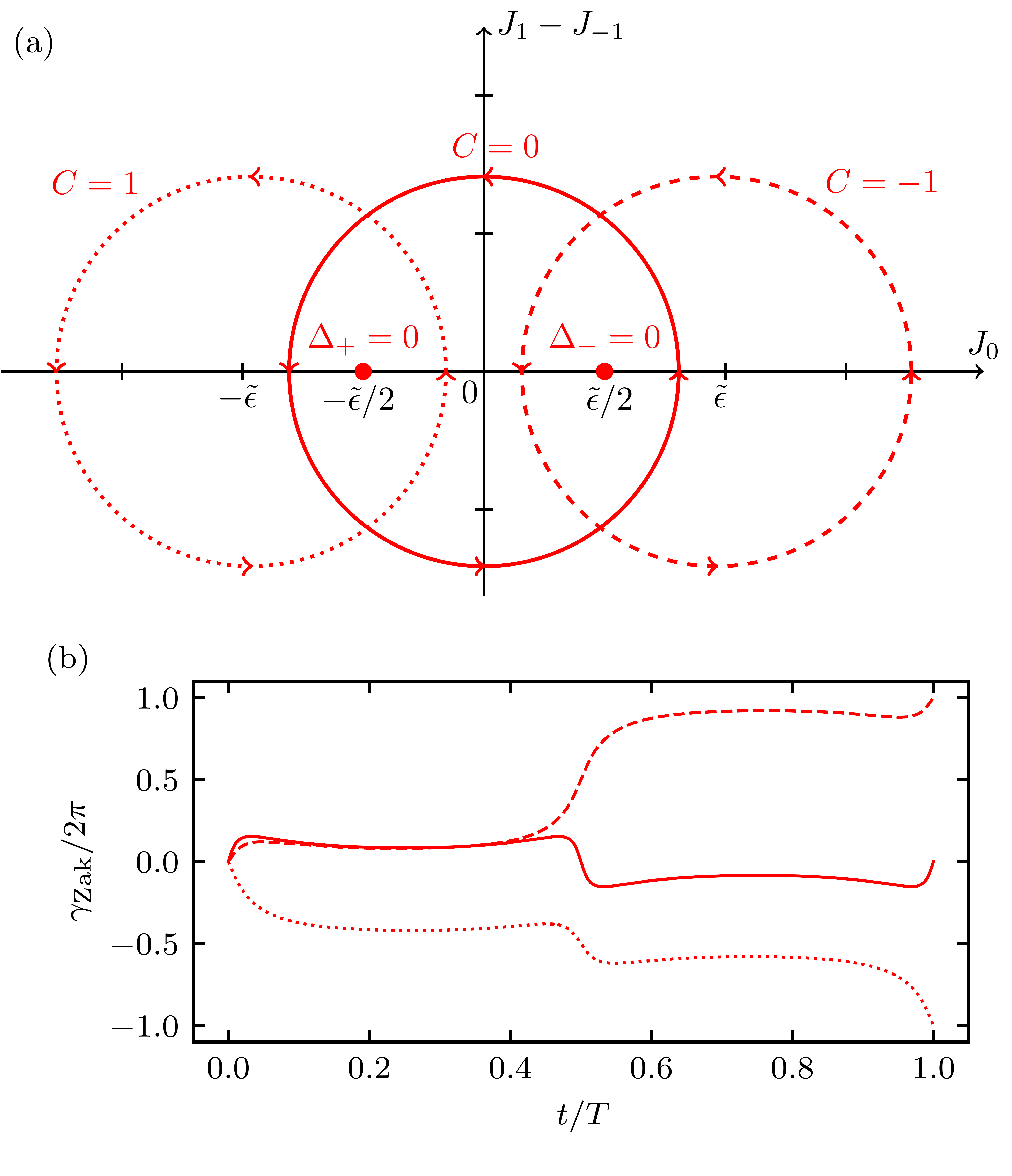}
\caption{Adiabatic pumping in the $J_0$ scheme. (a) Adiabatic pumping via modulation of $J_{\pm1}$ and $J_0$ when: both critical points $(\pm\tilde{\epsilon}/2, 0)$ are encircled (solid), the right point is encircled (dashed) or the left point is encircled (dotted). (b) Zak phase $\gamma_\mathrm{Zak}$ dependence on time $t$ for the three aforementioned trajectories.}
\label{fig:Zak_Combo-t0alpha}
\end{figure}

The red points in Fig.~\ref{fig:Zak_Combo-t0alpha}(a) denote
singular points in the $J_0$-$(J_1 - J_{-1})$ plane, and as above the circles plot different illustrative trajectories in this plane.
As shown by the solid circle, both singular points are enclosed when 
\begin{equation}
|J_{0,\mathrm{R}}| > \left|\frac{\tilde{\epsilon}}{2} \pm \bar{J}_0\right|\,.
\label{Condition-2}
\end{equation}
In this case, the detuning alternates for even versus odd $j$, so Eq.~\eqref{eq:Chern-singular points} sums to zero, giving Chern number $C=0$ and no adiabatic pumping.
On the other hand, when either condition in Eq.~\eqref{Condition-2} holds, the point at $(\pm\tilde{\epsilon}/2, 0)$ is encircled [the dotted and dashed circles in Fig.~\ref{fig:Zak_Combo-t0alpha}(a)], giving
$C=-1$ or $C=1$ respectively.
In these cases, adiabatic pumping gives a displacement of a single lattice constant $a$ (to the right or to the left): once again a scenario that is impossible for an uncoupled RM chain. 
Lastly when neither condition holds, no singular point is encircled and the system is again topologically trivial with $C=0$.

Figure~\ref{fig:Zak_Combo-t0alpha}(b) shows the Zak phase for the three trajectories in Fig.~\ref{fig:Zak_Combo-t0alpha}(a),
illustrating that the adiabatic pumping indeed yields displacement by 0 or $\pm a$.
We also evaluated the Wannier function centers by explicitly integrating the time dependent Schrodinger equation and found them to be in exact agreement with the displacements predicted by the Zak phase.

\section{Edge States}\label{sec:EdgeStates}

\begin{figure}[tbh]
\begin{centering}
\includegraphics[width=1\columnwidth]{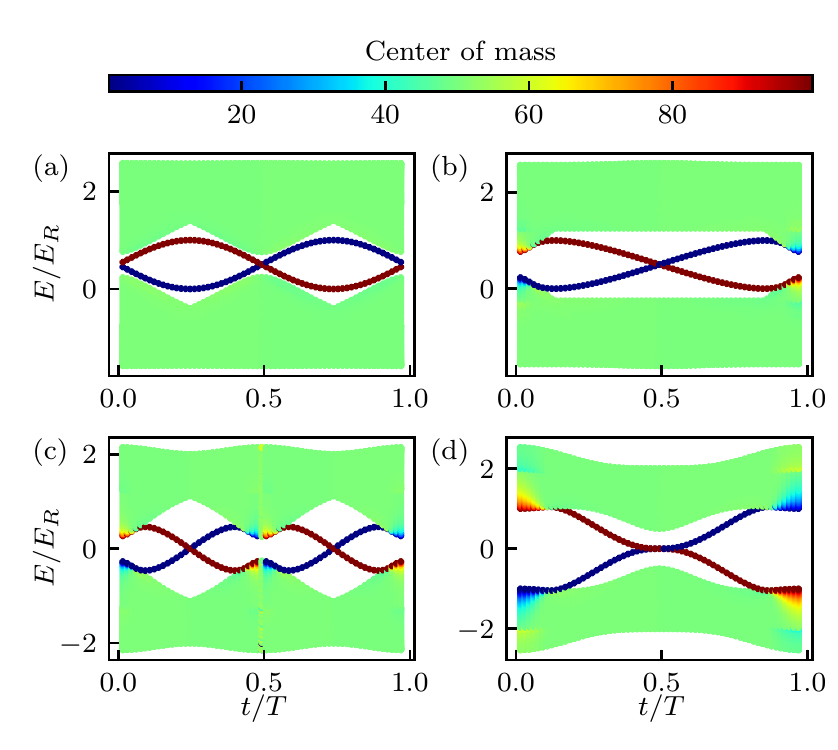}
\par\end{centering}
\caption{Energy spectrum of finite coupled RM chains with $L=100$ sites on each branch during adiabatic pumping cycle.
Color indicates center of mass position of the instantaneous eigenstates. (a) and (b) correspond to the $J_0$ modulation scheme, while (c) and (d) correspond to the $\epsilon$  scheme. (a) and (c) show the case with both special points encircled, while (b) and (d) show the case with only one point.}
\label{fig:edge1}
\end{figure}

The bulk-edge correspondence links the properties of bounded and un-bounded topological systems~\cite{Hatsugai1993,Halperin1982,hatsugai2022}.
This correspondence indicates that finite systems will acquire edge modes---residing in the bulk energy gap---and that the number of such modes will be equal to the infinite-system topological invariant.
We use this as an additional probe for the topology of coupled RM model, and turn our attention to a finite coupled RM chain with hard edges.

In Fig.~\ref{fig:edge1}, we plot the energy spectrum of eigenstates  (colored according to center of mass position) calculated for the $\epsilon$ and $J_0$ modulation schemes and for two different paths.
Exponentially localised edge states appear in the band-gap region, signaled by red/blue curves connecting bulk bands.
When these states become degenerate, the system obeys a chiral symmetry, making the static system Zak phase a robust topological invariant.
In this case these the degenerate states are the edge states predicted by the bulk-edge correspondence.

The movement of edge state between bands during a pump cycle marks the transport of mass from one side of the system to the other.
The first path encloses both special points, while the other only one. 
The exact trajectories are depicted in Fig.~\ref{fig:Zak_Combo-epsilon}(a) (solid and dashed curves) for the $\epsilon$ modulation scheme and Fig.~\ref{fig:Zak_Combo-t0alpha}(a) (solid and dotted curves) for the $J_0$ scheme.

We find a strict adherence to the bulk-edge correspondence: the number of edge states on each side of the system that flow from one band to another during each pump cycle is equal to the Chern number calculated in the previous section. 
Although Fig.~\ref{fig:edge1}~(b) and (d) qualitatively resemble the edge state flow of the conventional Rice-Mele model~\cite{Asboth_book}, the per-cycle displacement is reduced by half.
Special attention should be given to Fig.~\ref{fig:edge1} (a) which shows edge states shifting in energy between the bands but never entering into the bulk band, and therefore not contributing to charge pumping.
This further reinforces our conclusion, that the coupled RM chain described in this work is inherently different and cannot be explained by the superposition of two uncoupled RM chains.

\section{Conclusion and outlook}\label{sec:Discussion}

We described a lattice created by sequentially coupling internal atomic states using two photon Raman transitions; this resulted in independent adiabatic potentials for each of the $N$ dressed states.
We showed that introducing a static detuning couples together these adiabatic potentials into a single lattice  with $\lambda/(2N)$ periodicity.

We then studied the effects of time-modulated detuning to introduce effective tunneling matrix elements coupling the $s$- and $p$- bands., leading to a novel subwavelength optical lattice.
We interpreted this lattices as a pair of coupled RM chains described by a highly tunable two-leg ladder Hamiltonian with novel topological properties.
We showed that this lattice exhibits unusual behavior in terms of topological pumping and edge states. 
In the former case we showed that the added inter-chain tunneling enables simple pumping trajectories giving per-cycle displacements of 0, 1 or 2 unit cells; in contrast with the 0 or 2 cell displacements allowed for uncoupled RM chains.

The present work suggests several directions for future inquiry. 
Here we focused only on nearest neighbor detuning-induced tunneling, however,  more complicated lattice topologies can be created by exploiting the long-range tunneling induced by detuning in Eq.~\eqref{eq:delta_j-oscil-detun-1}.
Even without coupling $s$- and $p$-bands this may enable new ways to engineer locally flat bands~\cite{Sun2011a,Bergholtz2013} where interaction effects can dominate.

In addition, lattices in 2D and 3D can be created by going to larger spin systems; going beyond the suggestions in Ref.~\cite{Anderson2020,Zhou2022PRXQ} for creating conventional lattices, it is also possible to use coupled internal states to define a discretized torus rather than a closed loop in the space of coupled internal states.

The generic scheme described here is not limited to crystalline order.  
For example, a bichromatic subwavelength lattice can be created by adding Raman coupling with wavenumber incommensurate with $\kr$.
This would open up new possibilities to study localization phenomena with tunable single-particle mobility edges~\cite{Bloch2018PRL, DasSarma2017PRB}

Topological charge pumps obtain their robustness by fully filling a collection of Bloch bands.
One could investigate the more general case of geometric charge pumping~\cite{Lu2016} that lifts this constraint.
Given the flexibility of this lattice one might engineer the local Berry connection $A_k(k)$  to enhance the performance of geometric charge pumps, either for improved robustness, or even increased per-cycle displacement.

\begin{acknowledgments}
The authors thank T.~Andrijauskas and E.~Gvozdiovas for productive discussions, as well as A.~M.~Pi\~{n}eiro and M.~Zhao for carefully reading the manuscript.
Authors acknowledge support by the Lithuanian Research Council (Grant No. S-MIP-20-36).
This work was partially supported by the National Institute of Standards and Technology, the National Science Foundation through the Quantum Leap Challenge Institute for Robust Quantum Simulation (grant OMA-2120757), and the Air Force Office of Scientific Research office via the RAPSYDY in Q Multidisciplinary University Research Initiative (grant FA9550-22-1-0339).
\end{acknowledgments}

\appendix

\section*{Appendix}

\section{Zak phase and energy gaps}

\begin{figure}[tbh]
\includegraphics[width=0.5\textwidth]{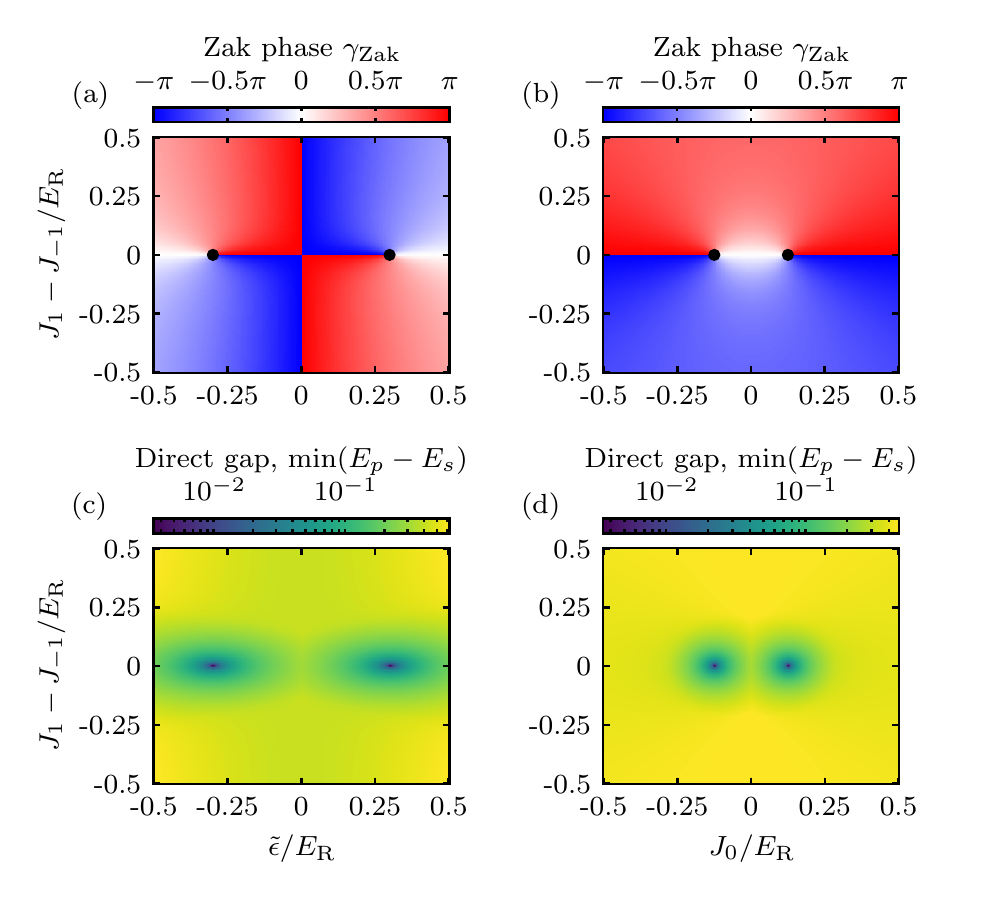}
\caption{Parameter regimes.
(a) and (b) Zak phase $\gamma_\mathrm{Zak}$.
(c) and (d) Direct energy gap $\mathrm{min} (E_p - E_s)$.
(a) and (c) correspond to the $\epsilon$ scheme, while (b) and (d) correspond to the $J_0$ scheme.
All figures are calculated for $J_\mathrm{avg}=0.1 E_\mathrm{R}$.
}
\label{fig:app:parameters}
\end{figure}

As shown in Fig.~\ref{fig:app:parameters}(a) and (b), to advance the Zak phase by $+2\pi$, one has to encircle either of the singular points in the same direction (counter-clockwise) for the $\epsilon$ scheme, but in opposite directions for the $J_0$ scheme (counter-clockwise for $\Delta_+ = 0$ and clockwise for $\Delta_- = 0$).
Thus, both critical points have the same topological charge for the $\epsilon$ case, but opposite charge in the $J_0$ case.

As one approaches the critical points in parameter space, the direct gaps grow smaller and smaller [see Fig.~\ref{fig:app:parameters}(c) and (d)].
Thus, if the encirclement radius is small, the adiabatic pumping period $T$ must be larger to satisfy the adiabaticity condition.

\clearpage

\bibliography{main}

\end{document}